\documentclass[preprint,12pt]{elsarticle}
\usepackage{graphicx}
\usepackage{graphics}
\usepackage{latexsym}
\usepackage{amsmath, amsthm, amssymb, wasysym}
\usepackage{xcolor}
\usepackage{soul}
\usepackage{cancel}
\usepackage{hyperref}
\usepackage[normalem]{ulem}

\usepackage{lineno}
\usepackage[english]{babel}

\journal{Physica A: Statistical Mechanics and its Applications}

\begin{document}
\begin{frontmatter}

\title{Spatial density, first passage times, and entropy\\of run and tumble particles in a bistable potential}

\author{R. K. Singh}
\ead{rksinghmp@gmail.com}
\address{Department of Physics, Banaras Hindu University, Varanasi 221005, India}

\author{Erez Aghion}
\ead{erez.aghion@louisiana.edu}
\address{Department of Physics, University of Louisiana at Lafayette, Lafayette, LA 70503, USA}

\begin{abstract}
We study the dynamics of a run and tumble particle (RTP) in a bistable potential, in contact with a heat bath.
We find the exact form of the steady state position distribution as well as its time-dependent form in Laplace space.
We use this result to study relaxation properties of the spatial probability density and find that it converges to the 
steady state from the center outwards, with the relaxation front moving ballistically in time.
Physically this ballistic behavior arises when observing relaxation properties
at the tails of the density, where the distance of the particle from the origin is larger than the length-scale describing the steady state of the RTP.
We also find the exact form of the mean first passage time, and the long-time behavior of the first passage time distribution. This allows us to coarse-grain the dynamics as a
dichotomous Markov process, a paradigm for a wide class of bistable processes. Coarse-graining, however, comes at the cost of a loss of information: while the RTP possesses a finite entropy production even at steady state, it identically vanishes
for its coarse-grained description.
\end{abstract}


\end{frontmatter}


\newcommand{\fr}{\frac}
\newcommand{\tl}{\tilde}
\newcommand{\dd}{\partial}
\newcommand{\lr}{\langle}
\newcommand{\rl}{\rangle}

\setlength\parindent{0pt}

\section{Introduction}
Bacterial motion in fluids is characterized by phases of \textit{runs} in which the bacteria travels in straight lines,
separated by \textit{tumbling} events marking changes in the direction of motion \cite{di2010bacterial,sokolov2010swimming,saragosti2011directional,
sidortsov2017role,taktikos2013motility,goral2022frustrated}. In a broader context, bacteria constitutes an example
of what is called an active particle, and in contrast to passive particles, they consume energy from the
environment for their motion \cite{bechinger2016active,ramaswamy2017active}. A wide variety of
processes ranging from stochastic dynamics in cells \cite{cates2012diffusive} to flocking \cite{toner2005hydrodynamics} exhibit active motion, thus requiring a broad set of specialized analytical techniques for their analysis~\cite{fodor2018statistical}.
Among these models, Run and Tumble Particles (RTPs) is a widely applicable 
model~\cite{solon2015active,cates2013active,elgeti2015run,ezhilan2015distribution,
angelani2017confined,angelani2023one,villa2020run,mori2020prl,mori2020universal}.

In one spatial dimension, the dynamics of a RTP is described by a pair of coupled Fokker-Planck equations,
one for each direction of movement. The positional probability density function 
(PDF) of an ensemble of free RTPs is given by a telegrapher’s 
equation~\cite{masoliver1996finite,porra1997telegrapher,weiss2002some}. Even though at times much longer
than the mean tumbling time this spatial PDF approaches a Gaussian, confinement is known to introduce
nontrivial effects. One of the prominent features of a RTP in a confining potential is that its steady state is
generally non-Boltzmann \cite{dhar2019run,sevilla2019stationary}. Moreover, for motion
in a harmonic potential, the steady state of the RTP is defined over a finite support \cite{dhar2019run,smith2022nonequilibrium},
in sharp contrast to a Brownian motion~\cite{risken1996fokker},  while in a periodic force field,
an RTP can exhibit a non-ergodic dynamics \cite{le2020velocity}.

A natural question arising from the above discussion is whether the steady state of a RTP in a confining potential
retains its finite support when the system is coupled to a heat bath.
The dynamics of an RTP coupled to a heat bath has been used to model active non-Markovian oscillations
\cite{tucci2022modeling}, study the dynamical properties in harmonic potential \cite{wexler2020dynamics,
garcia2021run,smith2022exact,frydel2022positing}, modeling cell division \cite{arcobi2023continuous}, and
to study the interaction of active particles \cite{le2021stationary}.
Below, we show that for a RTP moving
in a confining potential, in contact with heat bath, the steady state of the spatial PDF has  infinite support. Additionally, the coupling to a heat bath introduces some nontrivial features in the dynamics, affecting its relaxation and first passage properties. 

In a recent work~\cite{singh2025steady}, the transport properties of a RTP in contact with a heat bath, in a
piecewise linear confining potential, was studied. One of the key results of that work was finding the exact
form of the PDF of an RTP in Laplace space. In the current manuscript, we generalize  the previous study on monostable potentials
to bistable potentials and investigate both transport
and escape properties. We also study conditions under which the dynamics can be coarse-grained to a dichotomous
process.

Below, in Sec.~\ref{sec_dyn}, we describe the dynamics of a RTP in a bistable process,
followed by its steady state PDF in Sec. \ref{sec_px0}, and time-dependent solution in Sec. \ref{sec_time}. In
Sec.~\ref{sec_relax}, we study the relaxation toward the steady state, bridging the results of the two previous
sections. In Sec.~\ref{sec_mfpt}, we extend our study to first passage properties. We study the
properties of entropy in Sec.~\ref{sec_entropy} to examine the difference between the properties of RTPs and its related coarse-gained process.

\section{Run and tumble particle in a bistable potential}
\label{sec_dyn}
The dynamics of a RTP in a one-dimension, with a potential $U(x)$ and a constant-temperature heat
bath is described by the Langevin equation:
\begin{align}
\label{lang}
\frac{dx}{dt} = f(x(t)) +\sigma(t) + \eta(t).
\end{align}
Here, $f(x) = -dU(x)/dx$ is a deterministic force term, $\sigma(t)$ is active telegraphic (dichotomous) noise and $\eta(t)$ is thermal noise. The active noise $\sigma(t) \in \{-v_0,v_0\}$ has zero mean and correlation: $\lr \sigma(t) \sigma(s) \rl = v^2_0 e^{-2r|t-s|}$~\cite{bena2006dichotomous}, where $r$ is the rate of tumbling
and $v_0$ is the velocity of the RTP. The thermal noise is white and Gaussian, with zero mean and correlation: $\langle \eta(t) \eta(s) \rangle= 2\fr{k_BT}{\gamma}\delta(t-s)$. The constant $T$ is temperature, $\gamma$ is the friction coefficient, and $k_B$ is Boltzmann's constant.
In what follows, we rescale temperature with $\gamma/k_B$, which means $\lr \eta(t) \eta(s) \rl = 2T\delta(t-s)$.

Let $p_{\pm}(x,t)$ be the PDFs of an RTP at location $x$ at time $t$ with $\sigma(t)=\pm v_0$, respectively. 
The evolution of these PDFs is described by the coupled Fokker-Planck equations 
(FPEs)~\cite{risken1996fokker,singh2025steady}:
\begin{subequations}
\label{fpe}
\begin{align}
\dd_t p_+(x,t) &= -\dd_x (f(x)+v_0)p_+(x,t) - r p_+(x,t) + r p_-(x,t) + T \dd_{xx} p_+(x,t),\\
\dd_t p_-(x,t) &= -\dd_x (f(x)-v_0)p_-(x,t) + r p_+(x,t) - r p_-(x,t) + T \dd_{xx} p_-(x,t).
\end{align}
\end{subequations}  
In the above equations, the first and second order spatial derivatives describe respectively
the drift and coupling to the heat bath. The linear terms of $p_\pm(x,t)$ describe the tumbling events
as the RTP switches its direction of movement. Now, for the bistable potential, we use the piecewise linear function:
\begin{align}
\label{ux}
U(x) = \begin{cases}
b(x-a),~a\leq x,\\
-b(x-a),~0\leq x < a,\\
b(x+a),~-a\leq x < 0,\\
-b(x+a),~x\leq -a.
\end{cases}
\end{align}
The symmetry $U(-x) = U(x)$ means that~
$p_\pm(-x,t) = p_\mp(x,t)$.\\ 

Notably, the set of equations ~(\ref{fpe}a) and~(\ref{fpe}b) can also be
written in terms of probability currents $J_\pm(x,t) = (f(x(t))\pm v_0-T\dd_x)p_\pm(x,t)$ as:
\begin{align}
\dd_t p_\pm(x,t) = -\dd_x J_\pm(x,t) \pm g_0(x,t),
\end{align}
where $g_0(x,t) = -rp_+(x,t)+rp_-(x,t)$.
As we shall see below in the later sections, we will be using the probability currents $J_\pm(x,t)$
to describe entropy production and entropy for an RTP.

\subsection{Auxiliary variables}
To solve the FPEs in (\ref{fpe}), we define the following auxiliary
variables \cite{bier1993matching}:
\begin{subequations}
\label{auxl}
\begin{align}
P_0(x,t) &= p_+(x,t) + p_-(x,t),\\
P_1(x,t) &= \fr{d}{dx}P_0(x,t),\\
D_0(x,t) &= p_+(x,t) - p_-(x,t),\\
D_1(x,t) &= \fr{d}{dx}D_0(x,t).
\end{align}
\end{subequations}
The symmetry properties of $p_\pm(x,t)$ imply:
\begin{subequations}
\label{auxl_symm}
\begin{align}
P_0(-x,t) &= P_0(x,t),\\
D_0(-x,t) &= -D_0(x,t),\\
P_1(-x,t) &= -P_1(x,t),\\
D_1(-x,t) &= D_1(x,t).
\end{align}
\end{subequations}
Note that, although $P_0(x,t)$ and $D_0(x,t)$ are continuous, their derivatives $P_1(x,t)$ and $D_1(x,t)$,
exhibit jump discontinuities at the turning points of the potential $U(x)$. 
It is to be noted here that these jump discontinuities do not lead to any nonphysical artifacts as the magnitudes
of these jumps in either $P_1(x,t)$ or $D_1(x,t)$ at the turning points $x\in\{-a,0,a\}$ are consistent with the
changes in the slope of the potential $U(x)$ in the neighborhood of these turning points.
In what follows, using the symmetry of the potential we find the functions defined in Eqs.~(\ref{auxl})
only in the region $x\geq 0$, as the other side follows symmetrically.

\section{Steady state}
\label{sec_px0}
To find the shape of $p_\pm(x)$ at steady state, we solve:
\begin{subequations}
\label{fpe_ss}
\begin{align}
0 = &-\fr{d}{dx}(f(x)+v_0)p_+(x) - rp_+(x) + rp_-(x) + T\fr{d^2}{dx^2}p_+(x),\\
0 = &-\fr{d}{dx}(f(x)-v_0)p_-(x) + rp_+(x) - rp_-(x) + T\fr{d^2}{dx^2}p_-(x),
\end{align}
\end{subequations}
separately in two regions; first, at $a\leq x$, then in $0\leq x\leq a$. The steady state $P_0(x)$ then
follows by symmetry, that is, $P_0(-x) = P_0(x)$.

\subsection{Region $a\leq x$}
Using Eq.~\eqref{ux}, in this region, Eq.~\eqref{fpe_ss} in terms of the auxiliary variables in Eq.~(\ref{auxl}), reads:
\begin{align}
\label{mat3_ss}
&\fr{dP_0}{dx} = P_1,~\text{ and } \fr{d}{dx}
\begin{pmatrix}
P_1\\
D_0\\
D_1\\
\end{pmatrix}
= \underbrace{\begin{pmatrix}
-\fr{b}{T} & 0 & \fr{v_0}{T}\\
0 & 0 & 1\\
\fr{v_0}{T} & \fr{2r}{T} & -\fr{b}{T}
\end{pmatrix}}_{M_3(b)}
\begin{pmatrix}
P_1\\
D_0\\
D_1\\
\end{pmatrix}.
\end{align} 
To  solution to this set of equations is determined by the eigenvalues of the matrix $M_3(b)$,
following the characteristic equation:
\begin{align}
\label{eig3}
\lambda^3 + \fr{2b}{T}\lambda^2 - \Big(\fr{v^2_0}{T^2} + \fr{2r}{T}
- \fr{b^2}{T^2}\Big)\lambda - \fr{2rb}{T^2} = 0.
\end{align}
As the discriminant of the above cubic equation is positive (see \ref{AppR} for details), all the roots
of the characteristic equation are real, that is, the matrix $M_3(b)$ has real eigenvalues.
Furthermore,
\begin{subequations}
\begin{align}
&\text{tr}(M_3(b)) = -\fr{2b}{T} < 0,\text{ and }
\text{det}(M_3(b)) = \fr{2rb}{T^2} > 0,
\end{align}
\end{subequations}
with ``tr'' and ``det'' representing, respectively, the trace and determinant of $M_3(b)$. This implies that
two of the eigenvalues are negative and one is positive, with the three roots \cite{nickalls2006viete,zucker200892}:
\begin{align}
\label{lmb_k}
\lambda_j = -\fr{2b}{3T} + 2\sqrt{-\fr{q_1}{3}}\cos\Big[\fr{1}{3}\cos^{-1}\Big(\fr{3q_2}
{2q_1}\sqrt{-\fr{3}{q_1}}\Big)-\fr{2\pi j}{3}\Big],
\end{align}
for $j = 0, 1, 2$, with $q_1 = -\fr{v^2_0}{T^2}-\fr{b^2}{3T^2}-\fr{2r}{T}$ and
$q_2 = \fr{2bv^2_0}{3T^3}-\fr{2br}{3T^2}-\fr{2b^3}{27T^3}$.

Let $\lambda_1$ and $\lambda_2$ denote the two negative eigenvalues of $M_3(b)$, then the solution of the
matrix differential equation in (\ref{mat3_ss}) is:
\begin{align}
\begin{pmatrix}
P_1(x)\\
D_0(x)\\
D_1(x)\\
\end{pmatrix}
= \begin{pmatrix}
A_1\\
B_1\\
C_1\\
\end{pmatrix}e^{\lambda_1 x}
+ \begin{pmatrix}
A_2\\
B_2\\
C_2\\
\end{pmatrix}e^{\lambda_2 x}. 
\label{EqSolution}
\end{align}
The positive eigenvalue does not appear in this solution, since its contribution diverges at $x\rightarrow\infty$ and therefore it is non-physical. This means that its coefficient must be zero.  
The remaining two eigenvectors satisfy the equation:
\begin{align}
\begin{pmatrix}
-\fr{b}{T} & 0 & \fr{v_0}{T}\\
0 & 0 & 1\\
\fr{v_0}{T} & \fr{2r}{T} & -\fr{b}{T}
\end{pmatrix}
\begin{pmatrix}
A_k\\
B_k\\
C_k\\
\end{pmatrix} =
\lambda_k \begin{pmatrix}
A_k\\
B_k\\
C_k\\
\end{pmatrix},
\end{align}
From this equation and Eq.~\eqref{EqSolution}, we find:
\begin{subequations}
\label{sol_xgta}
\begin{align}
P_0(x) &= \fr{A_1}{\lambda_1}e^{\lambda_1 x} + \fr{A_2}{\lambda_2}e^{\lambda_2 x},\\
P_1(x) &= A_1 e^{\lambda_1 x} + A_2 e^{\lambda_2 x},\\
D_0(x) &= \fr{\lambda_1 T + b}{\lambda_1 v_0}A_1e^{\lambda_1 x} + \fr{\lambda_2 T + b}{\lambda_2 v_0}A_2
e^{\lambda_2 x},\\
D_1(x) &= \fr{\lambda_1 T + b}{v_0}A_1e^{\lambda_1 x} + \fr{\lambda_2 T + b}{v_0}A_2e^{\lambda_2 x}.
\end{align}
\end{subequations}
Here, $A_1$ and $A_2$ are coefficients which are calculated below.
It is to be noted here that the constant of integration appearing in $P_0(x)$ has been chosen to be zero under the condition that the PDF $P_0(x)$ approaches zero as $x\to \infty$. 
\subsection{Region $0\leq x < a$}
Proceeding analogously to the region $0\leq a\leq  x$, we have
\begin{subequations}
\label{sol_xlta}
\begin{align}
P_0(x) &= \fr{E_1}{\mu_1}e^{\mu_1 x} + \fr{E_2}{\mu_2}e^{\mu_2 x} + \fr{E_3}{\mu_3}e^{\mu_3 x},\\
P_1(x) &= E_1 e^{\mu_1 x} + E_2 e^{\mu_2 x} + E_3 e^{\mu_3 x},\\
D_0(x) &= \fr{\mu_1 T - b}{\mu_1 v_0}E_1e^{\mu_1 x} + \fr{\mu_2 T - b}{\mu_2 v_0}E_2
e^{\mu_2 x} + \fr{\mu_3 T - b}{\mu_3 v_0}E_3 e^{\mu_3 x},\\
D_1(x) &= \fr{\mu_1 T - b}{v_0}E_1e^{\mu_1 x} + \fr{\mu_2 T - b}{v_0}E_2
e^{\mu_2 x} + \fr{\mu_3 T - b}{v_0}E_3 e^{\mu_3 x},
\end{align}
\end{subequations}
with $\mu_1,~\mu_2,\text{and}~\mu_3$ being the three eigenvalues of $M_3(-b)$. Their values can be obtained
from Eqs.~(\ref{eig3}) and (\ref{lmb_k}) by replacing $b$ with $-b$. Furthermore,
the constant of integration in $P_0(x)$ has been chosen to be zero in the above set of equations.
The reason is that any such constant of integration within the region $0\leq x<a$ can be shifted to
the region $a\leq x$ following the continuity of $P_0(x)$ at the turning point of the potential. Moreover,
this constant of integration can be
eventually be made to vanish following the behavior of $P_0(x)$ in the limit $x\to\infty$.
The solution in Eq.~\eqref{sol_xlta} depends on three coefficients; $E_1, E_2$ and $E_3$. We will now find them.

\subsection{Evaluating the coefficients $A_1,~A_2,~E_1,~E_2$, and~$E_3$}
In order to estimate the coefficients in Eqs.~(\ref{sol_xgta}) and (\ref{sol_xlta}), we use the following relations:
\begin{enumerate}
\item Continuity of $P_0(x)$ at $x = a$, that is, $\lim_{x\to a^-}P_0(x) =
\lim_{x\to a^+}P_0(x)$.
\item Continuity of $D_0(x)$ at $x = a$, leading to $\lim_{x\to a^-}D_0(x) =
\lim_{x\to a^+}D_0(x)$.
\item Asymmetry of the flux; $D_0(-x) = -D_0(x)$, which leads to
$D_0(0) = 0$.
\item Taking difference of the two equations in Eq.~(\ref{fpe_ss}) results in:
$-\fr{d}{dx}f(x)D_0(x)-v_0\fr{d}{dx}P_0(x)-2rD_0(x)+T\fr{d}{dx}D_1(x) = 0$. Integrating this equation in the
neighborhood of $x = a$ provides the equation for jump discontinuity in $D_1(x)$: 
$\lim_{x\to a^+}D_1(x)-\lim_{x\to a^-}D_1(x) = -(2b/T)D_0(a)$.
\item And the normalization: $\int^\infty_0 dx~P_0(x) = 1/2$.
\end{enumerate}
Using Eqs.~(\ref{sol_xgta}) and (\ref{sol_xlta}) in conjunction with the above relations leads to the
following linear equation:
\begin{align}
\resizebox{\textwidth}{!}{$
\underbrace{\begin{pmatrix}
\fr{e^{\lambda_1 a}}{\lambda_1} & \fr{e^{\lambda_2 a}}{\lambda_2} & -\fr{e^{\mu_1 a}}{\mu_1}
& -\fr{e^{\mu_2 a}}{\mu_2} & -\fr{e^{\mu_3 a}}{\mu_3}\\
\fr{\lambda_1 T + b}{\lambda_1}e^{\lambda_1 a} & \fr{\lambda_2 T + b}{\lambda_2}e^{\lambda_2 a} &
-\fr{\mu_1 T - b}{\mu_1}e^{\mu_1 a} & -\fr{\mu_2 T - b}{\mu_2}e^{\mu_2 a} &-\fr{\mu_3 T - b}{\mu_3}e^{\mu_3 a}\\
0 & 0 & \fr{\mu_1 T - b}{\mu_1} & \fr{\mu_2 T - b}{\mu_2} & \fr{\mu_3 T - b}{\mu_3}\\
(\fr{2b}{\lambda_1}+T)(\lambda_1T+b)e^{\lambda_1 a} & (\fr{2b}{\lambda_2}+T)(\lambda_2T+b)e^{\lambda_2 a}
& -T(\mu_1T-b)e^{\mu_1 a} & -T(\mu_2T-b)e^{\mu_2 a} & -T(\mu_3T-b)e^{\mu_3 a}\\
-\fr{e^{\lambda_1 a}}{\lambda^2_1} & -\fr{e^{\lambda_2 a}}{\lambda^2_2} &\fr{e^{\mu_1 a}-1}{\mu^2_1}
& \fr{e^{\mu_2 a}-1}{\mu^2_2} & \fr{e^{\mu_3 a}-1}{\mu^2_3}\\
\end{pmatrix}}_{Q_5}
\begin{pmatrix}
A_1\\
A_2\\
E_1\\
E_2\\
E_3\\
\end{pmatrix}
= \begin{pmatrix}
0\\
0\\
0\\
0\\
1/2\\
\end{pmatrix}.
$}
\end{align}
The above equation can be solved using Cramer's rule \cite{arfken2011mathematical}, and the solution reads:
$A_1 = \Delta_1/\Delta,~A_2 = \Delta_2/\Delta,~E_1 = \Delta_3/\Delta,~E_2 = \Delta_4/\Delta,$ and
$E_3 = \Delta_5/\Delta$, where $\Delta$ is the determinant of matrix $Q_5$ and $\Delta_i,~i= 1,...,5$
are the determinants of the matrix obtained by replacing $i$-th column of $Q_5$ with $(0~0~0~0~1/2)^\top$ ($\top$ means transpose).

As a result, the steady state of the RTP in the region $0\leq x$ is:
\begin{align}
\label{p0x_st}
P_0(x) = \begin{cases}
\fr{1}{\Delta}\Big(\fr{\Delta_3}{\mu_1}e^{\mu_1 x} + \fr{\Delta_4}{\mu_2}e^{\mu_2x} +
\fr{\Delta_5}{\mu_3}e^{\mu_3 x}\Big),~0\leq x < a,\\
\fr{1}{\Delta}\Big(\fr{\Delta_1}{\lambda_1}e^{\lambda_1 x} + \fr{\Delta_2}{\lambda_2}e^{\lambda_2 x}\Big),~a\leq x.
\end{cases}
\end{align}
Moreover, the solution in the region $x\leq 0$ follows by symmetry as $P_0(-x) = P_0(x)$.
\begin{figure}[ht]
    \centering
    \includegraphics[width=1.0\textwidth]{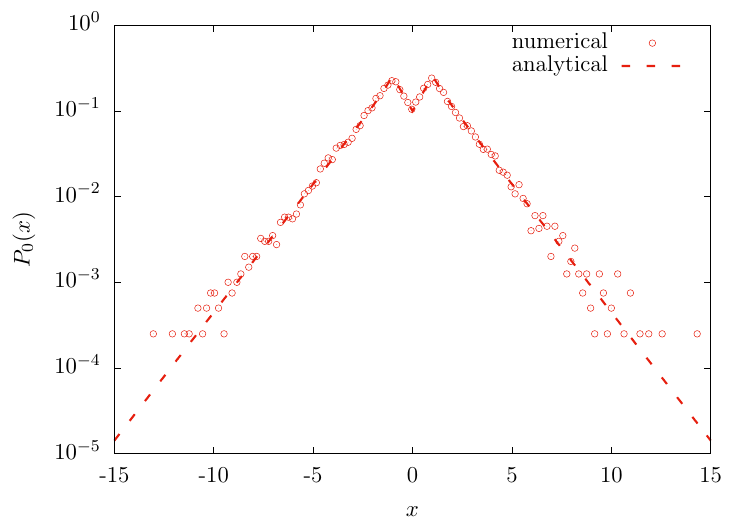}
    \caption{Steady state of a RTP in the bistable potential $U(x)$. The parameter values are: $a = 1,~b = 1,~r = 1,~v_0 = 
    1$, and $T = 1$. Numerical results are obtained from the solution of the Langevin equation (\ref{lang}) and the 
    analytical curves follow Eq.~(\ref{p0x_st}).
    } 
    \label{fig_ss}
\end{figure}
Fig.~\ref{fig_ss} presents simulation results, obtained by integrating over Eq.~\eqref{lang} numerically (see
\ref{App0} for details). These results are in excellent match with our analytical results describing
the steady state in (\ref{p0x_st}).

\section{Time-dependent solution}
\label{sec_time}
In order to solve the problem of calculating the time-dependent PDF of an RTP, we consider for simplicity
that the RTP starts at the origin with equal probability in velocity states $\pm v_0$, that is,
$p_\pm(x,0) = \delta(x)/2$.
To estimate the time-dependent shape of the PDFs in (\ref{fpe}), we employ Laplace transforms and use the 
auxiliary variables defined in Eq.~(\ref{auxl}).
The Laplace transform of any function $h(t)$ is $\tl{h}(s) = \int^\infty_0
dt~e^{-st}h(t)$. After transformation, Eq.~(\ref{fpe}) reads:
\begin{subequations}
\label{fpe_auxl}
\begin{align}
\label{fpe_auxl_p1}
s\tl{P}_0(x,s) - \delta(x) &= -\dd_x(f(x)\tl{P}_0(x,s)) - v_0\tl{D}_1(x,s) + T\dd_x\tl{P}_1(x,s),\\
\label{fpe_auxl_d1}
s\tl{D}_0(x,s) &= -\dd_x(f(x)\tl{D}_0(x,s)) - v_0\tl{P}_1(x,s) - 2r\tl{D}_0(x,s) + T\dd_x\tl{D}_1(x,s).
\end{align}
\end{subequations}
In conjunction with the definitions: $d\tl{P}_0(x,s)/dx = \tl{P}_1(x,s)$ and $d\tl{D}_0(x,s)/dx
= \tl{D}_1(x,s)$, we have a set of four coupled first order partial differential equations.
Analogous to the estimation of the steady state, we solve Eqs.~(\ref{fpe_auxl}) only
in the region $0\leq x$, as the solution for the remaining part $x\leq 0$ follows by symmetry.

\subsection{Region $a\leq x$}
In this region $f(x) = -b$, and as a result, Eq.~(\ref{fpe_auxl}) can be represented as the following matrix 
differential equation:
\begin{align}
\label{pxs_xgta}
\dd_x \begin{pmatrix}
\tl{P}_0\\
\tl{P}_1\\
\tl{D}_0\\
\tl{D}_1\\
\end{pmatrix} =
\underbrace{
\begin{pmatrix}
0 & 1 & 0 & 0\\
s/T & -b/T & 0 & v_0/T\\
0 & 0 & 0 & 1\\
0 & v_0/T & (2r+s)/T & -b/T\\
\end{pmatrix}
}_{M_4(b)}
\begin{pmatrix}
\tl{P}_0\\
\tl{P}_1\\
\tl{D}_0\\
\tl{D}_1\\
\end{pmatrix}.
\end{align}
Furthermore, for the matrix $M_4(b)$:
\begin{equation}
\text{tr}(M_4(b))= -2b/T <0,\text{ and } \text{det}(M_4(b))= s(2r+s)/T^2 >0.
\end{equation}
which implies that two of the eigenvalues of $M_4(b)$ are positive and the remaining two are
negative. This follows from the fact that in the limit $s\to 0$, one of the eigenvalues of
the matrix $M_4(b)$ approaches zero, while RTP approaches its steady state. Hence, the remaining three
eigenvalues should approach the eigenvalues of $M_3(b)$, which has two negative eigenvalues and one
positive eigenvalue.

Let $\lambda_1,~\lambda_2 < 0$ denote the two negative eigenvalues of $M_4(b)$, then:
\begin{subequations}
\label{px0s_vec_altx}
\begin{align}
\tl{P}_0(x,s) &= \fr{B_1}{\lambda_1}e^{\lambda_1 x} + \fr{B_2}{\lambda_2}e^{\lambda_2 x},\\
\tl{P}_1(x,s) &= B_1e^{\lambda_1 x} + B_2e^{\lambda_2 x},\\
\tl{D}_0(x,s) &= \fr{T}{v_0\lambda_1}\Big(\lambda_1+\fr{b}{T}-\fr{s}{T\lambda_1}\Big)B_1 e^{\lambda_1 x}
+ \fr{T}{v_0\lambda_2}\Big(\lambda_2+\fr{b}{T}-\fr{s}{T\lambda_2}\Big)B_2 e^{\lambda_2 x},\\
\tl{D}_1(x,s) &= \fr{T}{v_0}\Big(\lambda_1+\fr{b}{T}-\fr{s}{T\lambda_1}\Big)B_1 e^{\lambda_1 x}
+ \fr{T}{v_0}\Big(\lambda_2+\fr{b}{T}-\fr{s}{T\lambda_2}\Big)B_2 e^{\lambda_2 x}.
\end{align}
\end{subequations}
It is to be noted here that the explicit dependence of the eigenvalues $\lambda_1$ and $\lambda_2$ on
the Laplace parameter $s$ has been suppressed for brevity.
Once again, the coefficient before the terms resulting from the positive eigenvalues are set to zero, since they 
diverge at the large $x$ limit and are  thereby nonphysical.
And, analogous to the steady state solution, the constant of integration has been chosen to be zero
since $\tl{P}_0(x,s)$ vanishes when $x\to\infty$.

\subsection{Region $0< x <a$}
Similar to the previous region, in the region $0< x < a$, force on the RTP is $f(x) = b$, and as result,
the matrix differential equation is:
\begin{align}
\label{pxs_xlta}
\dd_x \begin{pmatrix}
\tl{P}_0\\
\tl{P}_1\\
\tl{D}_0\\
\tl{D}_1\\
\end{pmatrix} =
\underbrace{
\begin{pmatrix}
0 & 1 & 0 & 0\\
s/T & b/T & 0 & v_0/T\\
0 & 0 & 0 & 1\\
0 & v_0/T & (2r+s)/T & b/T\\
\end{pmatrix}
}_{M_4(-b)}
\begin{pmatrix}
\tl{P}_0\\
\tl{P}_1\\
\tl{D}_0\\
\tl{D}_1\\
\end{pmatrix}.
\end{align}
Let $\{\mu_i\}^4_{i=1}$ be the eigenvalues of $M_4(-b)$, then:
\begin{subequations}
\label{px0s_vec_xlta}
\begin{align}
\tl{P}_0(x,s) &= \fr{E_1}{\mu_1}e^{\mu_1 x} + \fr{E_2}{\mu_2}e^{\mu_2 x} + \fr{E_3}{\mu_3} e^{\mu_3 x}
+ \fr{E_4}{\mu_4}e^{\mu_4 x},\\
\tl{P}_1(x,s) &= E_1 e^{\mu_1 x} + E_2 e^{\mu_2 x} + E_3 e^{\mu_3 x} + E_4 e^{\mu_4 x},\\
\tl{D}_0(x,s) &= \fr{T}{v_0\mu_1}\Big(\mu_1-\fr{b}{T}-\fr{s}{T\mu_1}\Big)E_1 e^{\mu_1 x}
+ \fr{T}{v_0\mu_2}\Big(\mu_2-\fr{b}{T}-\fr{s}{T\mu_2}\Big)E_2 e^{\mu_2 x}\nonumber\\
&+ \fr{T}{v_0\mu_3}\Big(\mu_3-\fr{b}{T}-\fr{s}{T\mu_3}\Big)E_3 e^{\mu_3 x}
+ \fr{T}{v_0\mu_4}\Big(\mu_4-\fr{b}{T}-\fr{s}{T\mu_4}\Big)E_4 e^{\mu_4 x},\\
\tl{D}_1(x,s) &= \fr{T}{v_0}\Big(\mu_1-\fr{b}{T}-\fr{s}{T\mu_1}\Big)E_1 e^{\mu_1 x}
+ \fr{T}{v_0}\Big(\mu_2-\fr{b}{T}-\fr{s}{T\mu_2}\Big)E_2 e^{\mu_2 x}\nonumber\\
&+ \fr{T}{v_0}\Big(\mu_3-\fr{b}{T}-\fr{s}{T\mu_3}\Big)E_3 e^{\mu_3 x}
+ \fr{T}{v_0}\Big(\mu_4-\fr{b}{T}-\fr{s}{T\mu_4}\Big)E_4 e^{\mu_4 x},
\end{align}
\end{subequations}
wherein once again we have suppressed the explicit dependence of $\{\mu_i\}^4_{i=1}$ on $s$ for brevity.

\subsection{Evaluating the coefficients}
It is to be noted here that in both Eqs.~(\ref{px0s_vec_altx}) and (\ref{px0s_vec_xlta}), the eigenvalues
depend on the Laplace parameter $s$. In order to estimate the six coefficients, we shall make use of the
following relations:
\begin{enumerate}
\item Continuity of $\tl{P}_0(x,s)$ at $x = a$, that is, $\lim_{x\to a^-}\tl{P}_0(x,s) =
\lim_{x\to a^+}\tl{P}_0(x,s)$.
\item Continuity of $\tl{D}_0(x,s)$ at $x = a$, that is, $\lim_{x\to a^-}\tl{D}_0(x,s) =
\lim_{x\to a^+}\tl{D}_0(x,s)$.
\item Asymmetry of the flux, that is, $\tl{D}_0(-x,s) = -\tl{D}_0(x,s)$, which leads to
$\lim_{x\to 0^+} \tl{D}_0(x,s) = 0$.
\item Jump in the derivative of $\tl{P}_1(x,s)$ at $x = a$. This is obtained by integrating
Eq.~(\ref{fpe_auxl_p1}) in the neighborhood of $x = a$, which leads to: $\lim_{x\to a^+}\tl{P}_1(x,s) - \lim_{x\to a^-}
\tl{P}_1(x,s) = -(2b/T)\tl{P}_0(a,s)$.
\item Jump in the derivative of $\tl{D}_1(x,s)$ at $x = a$. This is obtained by integrating
Eq.~(\ref{fpe_auxl_d1}) in the neighborhood of $x = a$ and leads to: $\lim_{x\to a^+}\tl{D}_1(x,s) - \lim_{x\to a^-}
\tl{D}_1(x,s) = -(2b/T)\tl{D}_0(a,s)$.
\item And the normalization: $\int^\infty_0 dx~\tl{P}_0(x,s) = 1/2s$.
\end{enumerate}
Using the above relations we find that the coefficients $B_1,~B_2,~E_1,~E_2,~E_3$, and $E_4$ are
determined from the following set of linear equations:
\begin{align}
\hspace{-3em}\underbrace{
\begin{pmatrix}
\fr{e^{\lambda_1 a}}{\lambda_1} & \fr{e^{\lambda_2 a}}{\lambda_2} & -\fr{e^{\mu_1 a}}{\mu_1}
& -\fr{e^{\mu_2 a}}{\mu_2} & -\fr{e^{\mu_3 a}}{\mu_3} & -\fr{e^{\mu_4 a}}{\mu_4}\\
\nu_1 e^{\lambda_1 a} & \nu_2 e^{\lambda_2 a} & -\alpha_1 e^{\mu_1 a} & -\alpha_2 e^{\mu_2 a}
& -\alpha_3 e^{\mu_3 a} & -\alpha_4 e^{\mu_4 a}\\
0 & 0 & \alpha_1 & \alpha_2 & \alpha_3 & \alpha_4\\
(\fr{2b}{\lambda_1T}+1)e^{\lambda_1 a} & (\fr{2b}{\lambda_2T}+1)e^{\lambda_2 a}
& -e^{\mu_1 a} & -e^{\mu_2 a} & -e^{\mu_3 a} & -e^{\mu_4 a}\\
\nu_1\Big(\lambda_1+\fr{2b}{T}\Big) & \nu_2\Big(\lambda_2+\fr{2b}{T}\Big) & -\mu_1\alpha_1e^{\mu_1 a}
& -\mu_2\alpha_2e^{\mu_2 a} & -\mu_3\alpha_3e^{\mu_3 a} & -\mu_4\alpha_4e^{\mu_4 a}\\
-\fr{e^{\lambda_1 a}}{\lambda^2_1} & -\fr{e^{\lambda_2 a}}{\lambda^2_2} & \fr{e^{\mu_1 a}-1}{\mu^2_1}
& \fr{e^{\mu_2 a}-1}{\mu^2_2} & \fr{e^{\mu_3 a}-1}{\mu^2_3} & \fr{e^{\mu_4 a}-1}{\mu^2_4}\\
\end{pmatrix}
}_{Q_6}
\begin{pmatrix}
B_1\\
B_2\\
E_1\\
E_2\\
E_3\\
E_4\\
\end{pmatrix}
= \begin{pmatrix}
0\\
0\\
0\\
0\\
0\\
1/2s
\end{pmatrix},
\end{align}
where $\nu_i = \fr{T}{v_0\lambda_i}\Big(\lambda_i + \fr{b}{T} - \fr{s}{T\lambda_i}\Big)$ for $i \in \{1,2\}$
and $\alpha_i = \fr{T}{v_0\mu_i}\Big(\mu_i - \fr{b}{T} - \fr{s}{T\mu_i}\Big)$ for $i \in \{1,2,3,4\}$.
Using Cramer's rule \cite{arfken2011mathematical}, the solution of the above coupled linear equation reads:
$B_1 = \Delta_1/\Delta,~B_2 = \Delta_2/\Delta,~E_1 = \Delta_3/\Delta,~E_2 = \Delta_4/\Delta,~
E_3 = \Delta_5/\Delta$, and $E_4 = \Delta_6/\Delta$ where $\Delta$ is the determinant of matrix $Q_6$ and
$\Delta_i,~i\in \{1,...,6\}$ are the determinants of the matrix obtained by replacing $i$-th column of $Q_6$ with 
$(0~0~0~0~0~1/2s)^\top$, wherein $\top$ denotes transpose.
As a result, the PDF of the RTP in Laplace space in the region $0\leq x$ is:
\begin{align}
\label{p0x_ss}
\tl{P}_0(x,s) = \begin{cases}
\fr{1}{\Delta(s)}\Big(\fr{\Delta_3(s)}{\mu_1(s)}e^{\mu_1(s) x} + \fr{\Delta_4(s)}{\mu_2(s)}e^{\mu_2(s)x} +
\fr{\Delta_5(s)}{\mu_3(s)}e^{\mu_3(s) x} + \fr{\Delta_6(s)}{\mu_4(s)}e^{\mu_4(s) x}\Big),~0\leq x < a,\\
\fr{1}{\Delta(s)}\Big(\fr{\Delta_1(s)}{\lambda_1(s)}e^{\lambda_1(s) x} + \fr{\Delta_2(s)}{\lambda_2(s)}e^{\lambda_2(s) x}\Big),
~a\leq x,
\end{cases}
\end{align}
wherein we have now made the dependence on the Laplace parameter $s$ explicit.
In addition, the solution in the region $x\leq 0$ follows by symmetry as $\tl{P}_0(-x,s) = \tl{P}_0(x,s)$.

\section{Relaxation towards the steady state}
\label{sec_relax}
\begin{figure}[h]
    \centering
    \includegraphics[width=1.0\textwidth]{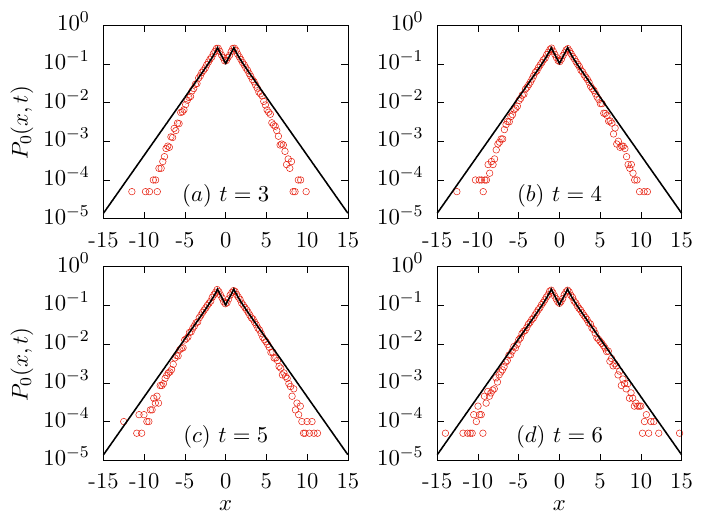}
    \caption{PDF of the RTP at different times. Red circles denote numerical estimates and the black
    solid line is the exact form of the steady state following Eq.~(\ref{p0x_st}). Parameter values are: $a = 1,~b = 1,
    ~r = 1,~v_0 = 1$, and $T = 1$.}
    \label{fig_relax}
\end{figure}
The PDF $P_0(x,t)$ of the RTP obtained from numerical simulation results, is presented at different times in Fig.~\ref{fig_relax}. The figure shows that $P_0(x,t)$ relaxes towards its steady state form given in Eq.~(\ref{p0x_st}) as time $t$ progresses.
This relaxation takes place from the center outwards, that is, the part close to the center relaxes first
and this relaxation front progresses outwards with time. Can we understand this relaxation
behavior by analyzing the Laplace transform $\tl{P}_0(x,s)$? 

The exact form of the PDF of the RTP in Laplace space, $\tl{P}_0(x,s)$, derived in Eq.~(\ref{p0x_ss}) is hard to 
invert, to provide us with the form of $P_0(x,t)$. However,
for a specific set of parameters, the Laplace transform can be inverted by employing numerical approximations for the 
eigenvalues and its coefficients. Hence, in this entire section we perform our study, for example,
with parameter values
$a = 1,~b = 1,~r = 1,~v_0 = 1$, and $T = 1$.
In order to proceed with our goal, we assume that the PDF $P_0(x,t)$
has already approximately relaxed within the spatial region $|x| \leq a$ (plausible in light of Fig.~\ref{fig_relax}),
and focus specifically on 
the region $a \leq |x|$, at long times. In other words, due to the Tauberian theorem~\cite{klafter2011first}, 
we study the properties of the Laplace transform $\tl{P}_0(x,s)$
for $a \leq |x|$ in the limit of small $s$, which corresponds to the long time limit.
From Eqs.~(\ref{px0s_vec_altx}) and (\ref{p0x_ss}), it follows that for $a \leq |x|$:
\begin{align}
\label{small_p0x}
\tl{P}_0(x,s) = \fr{B_1(s)}{\lambda_1(s)}e^{\lambda_1(s)|x|} + \fr{B_2(s)}{\lambda_2(s)}e^{\lambda_2(s)|x|},
\end{align}
with $\lambda_1(s),~\lambda_2(s) < 0$.
\begin{figure}[h]
    \centering
    \includegraphics[width=1.0\textwidth]{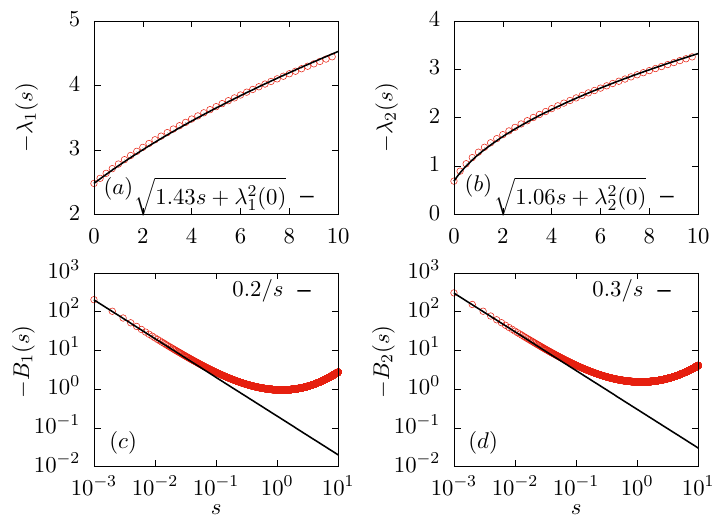}
    \caption{Small-$s$ behavior of the eigenvalues and coefficients defining the Laplace transform $\tl{P}_0(x,s)$.
    Parameter values are: $a = 1,~b = 1,~r = 1,~v_0 = 1$, and $T = 1$. }
    \label{fig_spectre}
\end{figure}
In Fig.~\ref{fig_spectre} we study the small-$s$ behavior of the eigenvalues $\{\lambda_1(s),\lambda_2(s)\}$,
and the coefficients $\{B_1(s),B_2(s)\}$ defining the Laplace transform $\tl{P}_0(x,s)$. Using fitting functions
for the eigenvalues and the coefficients in Eq.~(\ref{small_p0x}) we obtain:
\begin{align}
\label{P0xs_smalls}
\tl{P}_0(x,s) \stackrel{\text{small}~s}{\sim} \fr{0.2}{s\sqrt{l_1 s+\lambda^2_1(0)}}
e^{-\sqrt{l_1 s+\lambda^2_1(0)}x} + \fr{0.3}{s\sqrt{l_2 s+\lambda^2_2(0)}}
e^{-\sqrt{l_2 s+\lambda^2_2(0)}x},
\end{align}
where $l_1 \approx 1.43,~|\lambda_1(0)| \approx 2.48,~l_2 \approx 1.06$, and
$|\lambda_2(0)| \approx 0.69$ are fitting parameters. In addition,
we focus explicitly on the region $a \leq x$, as $x \leq -a$ follows by symmetry.
Now, the general form of the Laplace transforms in the above equation is:
\begin{align}
\tl{f}(s) = \fr{1}{s\sqrt{s+h}}e^{-\sqrt{s+h}y},
\end{align}
with $h\in\{\fr{\lambda^2_1(0)}{l_1},\fr{\lambda^2_2(0)}{l_2}\}$ and $y\in\{x\sqrt{l_1},x\sqrt{l_2}\}$. This
Laplace transform is straightforwardly inverted to obtain \cite{oberhettinger2012tables}:
\begin{align}
\label{intft00}
f(t) = \sqrt{\fr{t}{\pi}}\int^1_0 dw~\fr{e^{-t\phi(w)}}{\sqrt{w}},
\end{align}
with $\phi(w) = hw + y^2/4wt^2$. The long time limit of this integral is evaluated using the saddle-point
approximation \cite{arfken2011mathematical} and reads:
\begin{align}
\label{int_ft}
f(t) \stackrel{\text{large}~t}{\sim} \begin{cases}
\sqrt{\fr{2}{w_0|\phi''(w_0)|}}e^{-t\phi(w_0)},~\text{for}~w_0 < 1,\\
\sqrt{\fr{t}{\pi}}e^{-t\phi(1)},~\text{for}~1 \leq w_0.
\end{cases}
\end{align}
Here, $w_0= y/2t\sqrt{h}$ is the saddle point of $\phi(w)$, satisfying the relation $\phi'(w_0) = 0$.
Furthermore, each of the two Laplace transforms in Eq.~(\ref{P0xs_smalls}) have their respective
saddle points. Using Eq.~\eqref{int_ft} to invert the two Laplace transforms in Eq~.\eqref{P0xs_smalls} yields the 
long-time behavior of the PDF:
\begin{align}
\label{P0xt_larget}
\hspace{-2em}P_0(x,t) \stackrel{\text{large}~t}{\sim} \begin{cases}
\fr{0.2}{|\lambda_1(0)|}e^{-|\lambda_1(0)|x} + \fr{0.3}{|\lambda_2(0)|}e^{-|\lambda_2(0)|x},~a \leq x < \text{min}
\Big\{\fr{2|\lambda_1(0)|}{l_1}t,\fr{2|\lambda_2(0)|}{l_2}t\Big\},\\
0.2\sqrt{\fr{t}{\pi l_1}}e^{-t(\lambda^2_1(0)/l_1+l_1 x^2/4t^2)}
+ 0.3\sqrt{\fr{t}{\pi l_2}}e^{-t(\lambda^2_2(0)/l_1+l_2 x^2/4t^2)},~x > \text{max}
\Big\{\fr{2|\lambda_1(0)|}{l_1}t,\fr{2|\lambda_2(0)|}{l_2}t\Big\}.
\end{cases}
\end{align}
\begin{figure}[h]
    \centering
    \includegraphics[width=1.0\textwidth]{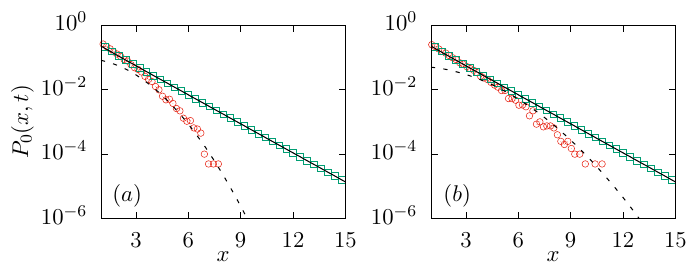}
    \caption{Relaxation behavior of the time-dependent PDF $P_0(x,t)$ at long times for (a) $t = 2$ and (b) $t = 4$.
    Red circles denote the numerically estimated $P_0(x,t)$ at time $t$, and green squares denote the steady state 
    form. Black solid line represents the relaxed part of $P_0(x,t)$ from Eq.~(\ref{P0xt_larget}) while the black 
    dashed lines represent the Gaussian fluctuations. Parameter values are: $a = 1,~b = 1,
    ~r = 1,~v_0 = 1$, and $T = 1$.}
    \label{fig_sdp}
\end{figure}
~\\
Equation~\eqref{P0xt_larget} implies that $P_0(x,t)$ approaches towards its steady state for large $t$.
The spatiotemporal boundary demarcating the region in which the PDF $P_0(x,t)$ has approximately achieved the
steady state is $a \leq |x| < \text{min} \Big\{\fr{2|\lambda_1(0)|}{l_1}t,\fr{2|\lambda_2(0)|}{l_2}t\Big\}$,
implying that the steady state front propagates ballistically. Physically, the
ballistic propagation of the relaxation front arises because we are observing the properties of the PDF
$P_0(x,t)$ on length-scales larger compared to the length-scale of relaxation, governed by $\text{max}\{|\lambda_1(0)|,
|\lambda_2(0)|\}$, at finite times $t$.
In addition, outside this region,
$P_0(x,t)$ is a mixture of Gaussians with their amplitudes decaying exponentially with time.
Numerical calculations in Fig.~\ref{fig_sdp} corroborate this assertion.

Now, using the symmetry properties of the time-dependent PDF, that is, $P_0(-x,t) = P_0(x,t)$,
we can rewrite Eq.~(\ref{P0xt_larget}) in an alternative manner as:
\begin{align}
|P_0(x,t)-P_0(x)| &\sim 0.2\sqrt{\fr{t}{\pi l_1}}e^{-t(\lambda^2_1(0)/l_1+l_1 x^2/4t^2)}
+ 0.3\sqrt{\fr{t}{\pi l_2}}e^{-t(\lambda^2_2(0)/l_1+l_2 x^2/4t^2)},\nonumber\\
&~\text{for}~|x| > \text{max}
\Big\{\fr{2|\lambda_1(0)|}{l_1}t,\fr{2|\lambda_2(0)|}{l_2}t\Big\},
\end{align}
at large times $t$. Moreover, the difference $|P_0(x,t)-P_0(x)|$ can be made arbitrarily small for large values
of $x$ and $t$, implying that the time-dependent PDF $P_0(x,t)$ remains only infinitesimally different
from its steady state limit $P_0(x)$.

We note that, while this method of inverting the Laplace transform $\tl{P}_0(x,s)$ can be easily
extended to other set of motion parameters, analysis with the chosen set provides us general insights 
the relaxation of the system to it steady state.

\section{Mean first passage time to the origin}
\label{sec_mfpt}
We study the mean first passage time (MFPT) of the RTP as it is an important quantifier of its search
efficiency. We find the MFPT using the backward Fokker-Planck equations for the survival 
probability $q_\pm(x,t)$ of the RTP starting at $x>0$, with an absorbing wall at $x=0$. These equations 
read~\cite{gueneau2025run}:
\begin{subequations}
\label{ddqxt}
\begin{align}
\dd_t q_+(x,t) &= (f(x)+v_0)\dd_x q_+(x,t) - rq_+(x,t) + rq_-(x,t) + T\dd_{xx}q_+(x,t),\\
\dd_t q_-(x,t) &= (f(x)-v_0)\dd_x q_-(x,t) + rq_+(x,t) - rq_-(x,t) + T\dd_{xx}q_-(x,t),
\end{align}
\end{subequations}
with absorbing boundary conditions $q_\pm(0,t) = 0$. The initial condition is 
$q_\pm(x,0) = 1/2$, assuming the RTP
starts with a state equally probable with both velocities $\pm v_0$. The first passage time distribution
(FPTD): $F(t) = -\fr{d}{dt}q(x,t)$, with $q(x,t) = q_+(x,t)+q_-(x,t)$ being the survival probability of
the RTP, while the MFPT to the origin 
is defined in terms of the survival probability as: $\tau_\pm(x) = \int^\infty_0 dt~q_\pm(x,t)$.
Since $q_+(x,t)=q_-(x,t)=1/2,$ integrating the
above backward equations with respect to  time, $t$, yields:
\begin{subequations}
\label{ddmfpt}
\begin{align}
-\fr{1}{2} &= (f(x)+v_0)\fr{d}{dx}\tau_+ - r\tau_+ + r\tau_- + T\fr{d^2}{dx^2}\tau_+,\\
-\fr{1}{2} &= (f(x)-v_0)\fr{d}{dx}\tau_- + r\tau_+ - r\tau_- + T\fr{d^2}{dx^2}\tau_-.
\end{align}
\end{subequations}
Now, in the region $0\leq x$ the potential in Eq.~(\ref{ux}) is: $U(x) = b|x-a|$, hence $f(x)= -b~\text{sign}(x-a)$.
As a result, the survival probability
$q_\pm(x,t)$ is continuous, while its spatial derivative $\dd_x q_\pm(x,t)$ exhibits a discontinuity at the turning
point $x = a$. Notwithstanding, the MFPTs $\tau_\pm(x)$ satisfy the same boundary conditions as $q_\pm(x,t)$, except
at the turning point, the derivatives $\fr{d}{dx}\tau_\pm(x)$ are continuous, as the MFPTs are the integral of 
$q_\pm(x,t)$. In order to solve Eq.~(\ref{ddmfpt}) we introduce the auxiliary variables $S_0 = \tau_+ + \tau_-,~dS_0/
dx = S_1,~D_0 = \tau_+ - \tau_-$, and $dD_0/dx = D_1$, and again solve the problem in two spatial regimes,
that is, $x\leq a$ and $a\leq x$.

\subsection{Mean first passage time from the region $a\leq x$} 
In this region, Eq.~(\ref{ddmfpt}) in terms of the auxiliary varibles reads:
\begin{align}
\fr{d}{dx}\begin{pmatrix}
S_1\\
D_0\\
D_1\\
\end{pmatrix} =
\underbrace{
\begin{pmatrix}
b/T & 0 & -v_0/T\\
0 & 0 & 1\\
-v_0/T & 2r/T & b/T\\
\end{pmatrix}
}_{Q_3(b)}
\begin{pmatrix}
S_1\\
D_0\\
D_1\\
\end{pmatrix}
+ \begin{pmatrix}
-1/T\\
0\\
0\\
\end{pmatrix}.
\end{align}
Let $\alpha$ denote the lone negative eigenvalue of $Q_3(b)$, then the physically consistent solutions
are:
\begin{subequations}
\begin{align}
S_1(x) &= Ae^{\alpha x} + \fr{1}{b},\\
D_0(x) &= \fr{AT}{\alpha v_0}\Big(\fr{b}{T}-\alpha\Big)e^{\alpha x} + \fr{v_0}{2br},\\
D_1(x) &= \fr{AT}{v_0}\Big(\fr{b}{T}-\alpha\Big)e^{\alpha x}.
\end{align}
\end{subequations}
As a result, $dS_0/dx = S_1$ implies
\begin{align}
\label{s0x_mfpt}
S_0(x) = \fr{A}{\alpha}e^{\alpha x} + \fr{x}{b} + C,
\end{align}
with $C$ being the constant of integration.

\subsection{Mean first passage time from the region $0\leq x < a$}
In this region, the solution of Eqs.~(\ref{ddmfpt}) is:
\begin{subequations}
\begin{align}
S_0(x) &= \fr{B_1}{\eta_1}e^{\eta_1 x} + \fr{B_2}{\eta_2}e^{\eta_2 x} + \fr{B_3}{\eta_3}e^{\eta_3 x} - \fr{x}{b} + E,\\
S_1(x) &= B_1 e^{\eta_1 x} + B_2 e^{\eta_2 x} + B_3 e^{\eta_3 x}-\fr{1}{b},\\
D_0(x) &= -\fr{T B_1\Big(\eta_1 + \fr{b}{T}\Big)}{v_0\eta_1} e^{\eta_1 x}
-\fr{TB_2\Big(\eta_2 + \fr{b}{T}\Big)}{v_0\eta_2} e^{\eta_2 x}
-\fr{T B_3\Big(\eta_3 + \fr{b}{T}\Big)}{v_0\eta_3} e^{\eta_3 x} - \fr{v_0}{2br},\\
D_1(x) &= -\fr{T\Big(\eta_1 + \fr{b}{T}\Big)B_1 }{v_0}e^{\eta_1 x}
-\fr{T\Big(\eta_2 + \fr{b}{T}\Big)B_2}{v_0} e^{\eta_2 x}
-\fr{T\Big(\eta_3 + \fr{b}{T}\Big)B_3}{v_0} e^{\eta_3 x},
\end{align}
\end{subequations}
wherein $\{\eta_1,\eta_2,\eta_3\}$ are the eigenvalues of $Q_3(-b)$ and $E$ being the constant of
integration. Their values can be determined
from Eqs.~(\ref{eig3}) and (\ref{lmb_k}) by replacing $v_0$ by $-v_0$ and $-b$ with $b$.

\subsection{Evaluating the coefficients}
In order to evaluate the coefficients $A,C,E; B_1,B_2,B_3$ we make use of the following conditions:
\begin{enumerate}
\item Absorbing boundary condition at the origin: $S_0(0) = 0$ and $D_0(0) = 0$,
\item Continuity at the turning point of the potential: $\lim_{x\to a^-}S_0(x) = \lim_{x\to a^+}S_0(x)$ and
$\lim_{x\to a^-}D_0(x) = \lim_{x\to a^+}D_0(x)$,
\item Continuity of the derivative at the turning point: $\lim_{x\to a^-}S_1(x) = \lim_{x\to a^+}S_1(x)$
and $\lim_{x\to a^-}D_1(x) = \lim_{x\to a^+}D_1(x)$.
\end{enumerate}
Using these conditions we find the following set of linear equations connecting all the coefficients describing
the MFPTs:
\begin{align}
\underbrace{
\begin{pmatrix}
0 & 0 & 1 & \fr{1}{\eta_1} & \fr{1}{\eta_2} & \fr{1}{\eta_3}\\
0 & 0 & 0 & \fr{1}{\eta_1}\Big(\eta_1 + \fr{b}{T}\Big) & \fr{1}{\eta_2}\Big(\eta_2 + \fr{b}{T}\Big)
& \fr{1}{\eta_3}\Big(\eta_3 + \fr{b}{T}\Big)\\
-\fr{e^{\alpha a}}{\alpha} & -1 & 1 & \fr{e^{\eta_1 a}}{\eta_1} & \fr{e^{\eta_2 a}}{\eta_2} & \fr{e^{\eta_3 a}}{\eta_3}\\
\fr{e^{\alpha a}}{\alpha}\Big(\fr{b}{T}-\alpha\Big) & 0 & 0 & \fr{e^{\eta_1 a}}{\eta_1}\Big(\eta_1+\fr{b}{T}\Big)
& \fr{e^{\eta_2 a}}{\eta_2}\Big(\eta_2+\fr{b}{T}\Big) & \fr{e^{\eta_3 a}}{\eta_3}\Big(\eta_3+\fr{b}{T}\Big)\\
-e^{\alpha a} & 0 & 0 & e^{\eta_1 a} & e^{\eta_2 a} & e^{\eta_3 a}\\
e^{\alpha a}\Big(\fr{b}{T}-\alpha\Big) & 0 & 0 & e^{\eta_1 a}\Big(\eta_1+\fr{b}{T}\Big) &
e^{\eta_2 a}\Big(\eta_2+\fr{b}{T}\Big) & e^{\eta_3 a}\Big(\eta_3+\fr{b}{T}\Big)
\end{pmatrix}
}_{R_6}
\begin{pmatrix}
A\\
C\\
E\\
B_1\\
B_2\\
B_3\\
\end{pmatrix} =
\begin{pmatrix}
0\\
-v^2_0/2brT\\
2a/b\\
-v^2_0/brT\\
2/b\\
0
\end{pmatrix}.
\end{align}
The solution of the above equation is obtained using Cramer's rule \cite{arfken2011mathematical}:
$A = \Delta_1/\Delta,~C = \Delta_2/\Delta,~E = \Delta_3/\Delta,~B_1 = \Delta_4/\Delta,~B_2 = \Delta_5/\Delta,$ and
$B_3 = \Delta_6/\Delta$. Here $\Delta = \text{det} R_6$ and $\Delta_i$ are the determinants of the
matrix obtained by replacing the $i$-th column of $R_6$ with the column vector $(0~ -v^2_0/2brT~ 2a/b~ -v^2_0/brT~
2/b~0)^\top$.

As a result, the MFPT of the RTP to go from a location $x>0$ to cross the energy barrier located at $x=0$,
following Eq.~(\ref{s0x_mfpt}) is:
\begin{align}
\label{mfpt00}
\tau(x) = \fr{\Delta_1 e^{\alpha x} + \alpha\Delta_2}{\alpha\Delta} + \fr{x}{b},~a\leq x.
\end{align}

\subsection{Distribution of first passage times}
\begin{figure}[h]
    \centering
    \includegraphics[width=1.0\textwidth]{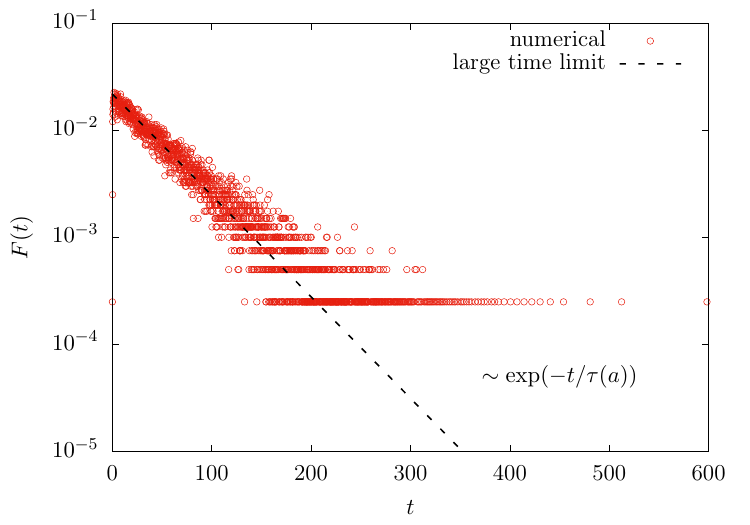}
    \caption{First passage time distribution $F(t)$ of a RTP starting at the energy minima at $x = a$ and crossing
    the energy barrier at the origin (red circles). Parameter values are: $r = 1,~a = 1,~b = 1,~T = 1/5$, and $v_0 = 
    1/4$. The dashed line represents the long time limit of $F(t)$ with the time-scale $\tau(a)$ obtained from 
    Eq.~(\ref{mfpt00}).}
    \label{fig_fptd}
\end{figure}
We find the FPTD $F(t)$ of an RTP starting at the energy minima
on the positive $x$ side and escaping through the absorbing wall at the origin to the negative $x$ region.
For this purpose, we solve an example where we choose the parameter values as: $r = 1,~a = 1,~b = 1,~T = 1/5$, and $v_0 = 1/4$.
We estimate $F(t)$ via the numerical solution of the Langevin equation (\ref{lang}).
Numerical estimates further suggest that $\tau(a) \approx 50.82$ 
while the analytical estimate obtained from
Eq.~(\ref{mfpt00}) reads $\tau(a) \approx 45.88$, in fairly good agreement with the numerical
results.

Furthermore,
the result in Fig.~\ref{fig_fptd} also indicates: $F(t) \stackrel{\text{large}~t}{\sim} \exp(-t/\tau(a))$,
implying that at long times the FPTD exhibits exponential decay. This is an important result showing that in the limit
of large $U(0)/T$ (its value is 5 in Fig.~\ref{fig_fptd}), the FPTD can be approximated by its long time limit.

\section{Entropy and coarse-graining}
\label{sec_entropy}
As RTPs constitute a classic example of out-of-equilibrium systems, studying its thermodynamic properties like
entropy becomes physically important \cite{padmanabha2023fluctuations}, and hence, we study the 
entropy of an
RTP in the bistable potential. The entropy of a RTP is defined as:
$S(t) = -\int^\infty_{-\infty} dx~p_+(x,t)\ln p_+(x,t)-\int^\infty_{-\infty} dx~p_-(x,t)\ln p_-(x,t)$.
The total entropy can be rearranged and written 
as the sum of entropy production and entropy flow\cite{entropy}:
\begin{align}
\label{stsf}
\fr{d}{dt}S(t) &= \underbrace{\fr{1}{T}\Big(\int dx~\fr{J^2_+}{p_+} + \int dx~\fr{J^2_-}{p_-}\Big) + r\int dx~(p_+-p_-)
\ln\fr{p_+}{p_-}}_{\text{entropy production}~dS_i/dt}\nonumber\\
&-\underbrace{\Big(\fr{1}{T}\int dx~f(x)(J_++J_-)+\fr{v_0}{T}\int dx~(J_+-J_-)\Big)}_{\text{entropy flow}~dS_e/dt}.
\end{align}
The non-negative nature of entropy production connects it to heat dissipated to the medium \cite{Seifert}.
Now an  estimation of either
of the two terms, $\fr{dS_i}{dt}$ or $\fr{dS_e}{dt}$, as a function of time $t$ is a nontrivial problem for the reason
that an evaluation of $p_\pm(x,t)$ is hard. Even though we derive exact forms of $\tl{p}_\pm(x,s)$ (see 
Sec.~\ref{sec_time}), 
inverting the resulting Laplace transforms for a given set of parameters is a problem in itself and is, for now, left 
for future investigation. Therefore, we calculate the entropy analytically at steady state and use it to calculate
entropy production in terms of entropy flow.

\subsection{Evaluating the entropy at steady state}
At steady state, the entropy of the system is maximal, and its value is:
\begin{align}
\label{ent_ss}
\lim_{t\to\infty}S(t) &= -\int^\infty_0 dx~(P_0(x)-D_0(x))\ln\fr{P_0(x)-D_0(x)}{2}\nonumber\\
&-\int^\infty_0 dx~(P_0(x)+D_0(x))\ln\fr{P_0(x)+D_0(x)}{2},
\end{align}
wherein we have used the symmetry properties of $P_0(x)$ and $D_0(x)$ described in
Eq.~(\ref{auxl_symm}).
The values of the above integrals can be estimated from the steady state solution in Eqs.~(\ref{sol_xgta}) and
(\ref{sol_xlta}), that is:
\begin{align}
\label{p0pmd0}
P_0(x) \pm D_0(x) = \begin{cases}
\fr{1}{\Delta}\sum^3_{i=1}\Big(1\pm\fr{\mu_iT-b}{v_0}\Big)\fr{\Delta_{i+2}}{\mu_i}e^{\mu_i x},~0\leq x< a,\\
\fr{1}{\Delta}\sum^2_{i=1}\Big(1\pm\fr{\lambda_i T + b}{v_0}\Big)
\fr{\Delta_i}{\lambda_i}e^{\lambda_i x},~a\leq x.
\end{cases}
\end{align}
Since the rate of change of the total entropy vanishes at the steady state, that is, $dS/dt = 0$, the entropy 
production in this state equals entropy flow. This implies that the entropy production $dS_i/dt$ at steady
state is:
\begin{align}
\label{dsedt00}
\fr{dS_i}{dt} &= \fr{1}{T}\int^\infty_{-\infty} dx~f(x)(J_+(x)+J_-(x))+\fr{v_0}{T}
\int^\infty_{-\infty} dx~(J_+(x)-J_-(x)),\nonumber\\
&= \fr{v_0}{T} \int^\infty_{-\infty} dx~(J_+(x)-J_-(x)).
\end{align}
Here, the equality in the second line appears since the integral involving the total current $J_++J_-$ vanishes identically at steady state.
Now, expressing the currents $J_\pm(x)$ in terms of the auxiliary variables in Eq.~\eqref{auxl}, we find:
\begin{align}
\label{jpmjm}
J_+(x)-J_-(x) &= f(x)D_0(x) + v_0P_0(x) - TD_1(x).
\end{align}
Using Eqs.~(\ref{sol_xgta}) and
(\ref{sol_xlta}) from Sec.~\ref{sec_px0}, and the determinants defined in Eq.~(\ref{p0x_st}), the entropy production of the RTP can be 
estimated in the steady state.
The advantage of this approach, as opposed to directly using the expression for
entropy production (see Eq.~(\ref{stsf})), lies in simplification of the integrands expressed in terms of
sums and differences of probability currents.
As a result, the entropy production at steady state is:
\begin{align}
\fr{dS_i}{dt} &= \fr{v^2_0}{T} + \fr{2bv_0}{T}\Big(\int^a_0 dx~D_0(x) - \int^\infty_a dx~D_0(x)\Big),
\end{align}
wherein we have made use of the symmetry properties of the auxiliary variables $D_0$ and $D_1$.
A non-zero value of the entropy production $dS_i/dt$ at the steady state implies the presence of nonvanishing
fluctuations as observed in Fig.~\ref{fig_traj}. Now the symmetry of the bistable potential $U(x)$ implies
equal occupancy in both the wells. However, for barrier heights large compared to the temperature, that is,
$U(0)/T\gg 1$, 
the time spent in the energy wells is much longer than the time required to make a transit from one well 
to another (see Fig.~\ref{fig_traj}). In such a scenario, the dynamics of the RTP in the bistable potential
can be coarse-grained and studied as effectively a dichotomous process. Such a coarse-grained description is
useful as many different class of systems undergoing bistable dynamics can be mapped to it
\cite{bena2006dichotomous}.

\subsection{Evaluating the entropy of coarse-grained dynamics}
\begin{figure}[h]
    \centering
    \includegraphics[width=1.0\textwidth]{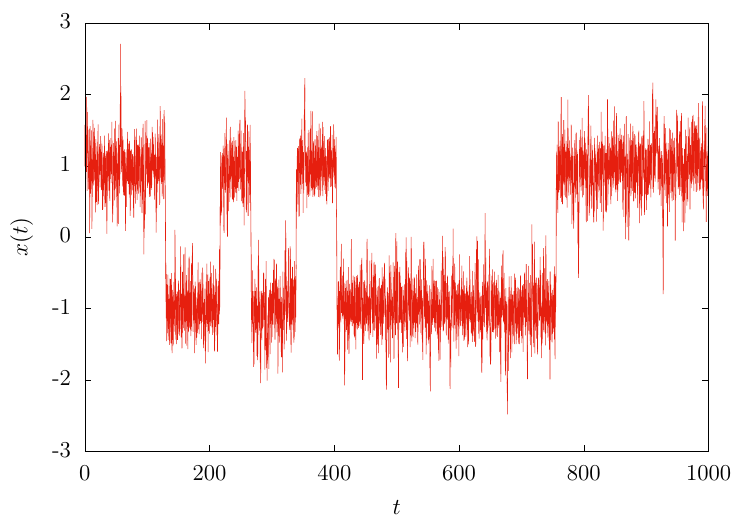}
    \caption{Typical trajectory of the RTP in the bistable potential $U(x)$. Parameter values are: $r = 1,~a = 1,~b = 1,~T = 1/5$, 
    and $v_0 = 1/4$.}
    \label{fig_traj}
\end{figure}
As discussed above, for large $U(0)/T$, the RTP tends to spend maximal time in either of the energy wells, with the
mean residence time (MRT) given in Eq.~(\ref{mfpt00}).
The result in Fig.~\ref{fig_traj} also shows that the dynamics of a RTP in the bistable potential
$U(x)$ can be coarse-grained to represent the dynamics of a particle jumping between states $\{-1,1\}$ with
exponentially distributed residence times with parameter $\tau\equiv\tau(a)$.

At a coarse-grained level, the dynamics of the RTP in the bistable potential can be described by a dichotomous
process. Furthermore, the exponentially distributed residence times imply that the coarse-grained process is also
Markovian. If $\rho_\pm(t)$ denotes the probability of the RTP to be in the potential well located at $\pm a$, then
\cite{bena2006dichotomous,toral2014stochastic}:
\begin{align}
\label{two_st}
\fr{d}{dt}\begin{pmatrix}
\rho_+(t)\\
\rho_-(t)\\
\end{pmatrix}
= \begin{pmatrix}
-1/\tau & 1/\tau\\
1/\tau & -1/\tau\\
\end{pmatrix}
\begin{pmatrix}
\rho_+(t)\\
\rho_-(t)\\
\end{pmatrix}.
\end{align}
Moreover, if the RTP initially starts at $x = a$, that is, $\rho_+(0) = 1$, then 
\cite{bena2006dichotomous,toral2014stochastic}:
\begin{subequations}
\label{rho_t}
\begin{align}
\rho_+(t) &= \fr{1}{2}(1+e^{-2t/\tau}),\\
\rho_-(t) &= \fr{1}{2}(1-e^{-2t/\tau}),
\end{align}   
\end{subequations}
wherein we refer $\tau(a)$ as $\tau$ for brevity. It is evident from the above equations that
$\rho_\pm(t) \stackrel{t/\tau\gg 1}{\sim} 1/2$, that is, both the energy wells are equally resided at times
much longer than the MRT $\tau$.

\begin{figure}
    \centering
    \includegraphics[width=1.0\textwidth]{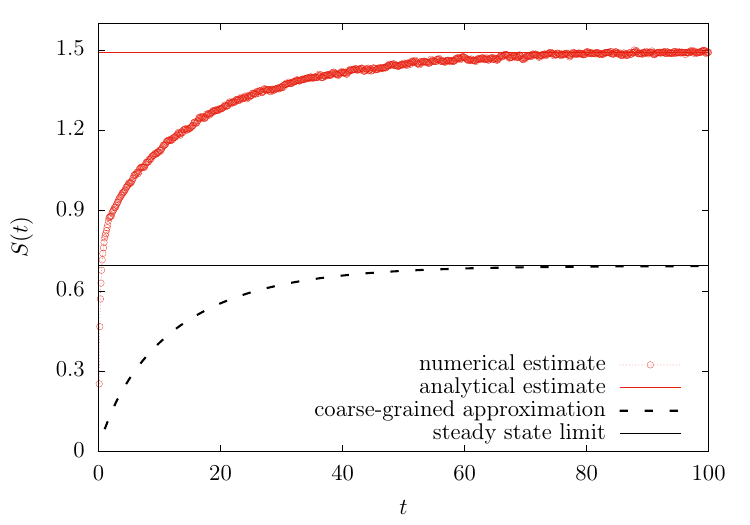}
    \caption{Numerical estimation of entropy $S(t)$ of the RTP as a function of time $t$
    (red circles), and its comparison with the analytical estimate obtained from the steady state solution (red solid 
    line, from Eq.~(\ref{ent_ss})). For comparison,
    the coarse-grained approximation is also shown (black dashed line). Black solid line represents the steady
    state limit of the coarse-grained dynamics and equals $\ln 2$. Parameter values are:
    $r = 1,~a = 1,~b = 1,~T = 1/5$, and $v_0 = 1/4$. }
    \label{fig_entropy}
\end{figure}
The coarse-grained distributions in Eq.~(\ref{rho_t}) allow us to estimate the average information content of the 
system, namely the coarse-grained entropy;~$S^{cg}(t)$:
\begin{align}
\label{cgst}
S^{cg}(t) = -\lr \ln\rho(t) \rl = \ln 2 - \sum_{i\in\{+,-\}}\rho_i\ln(2\rho_i(t)),
\end{align}
implying that $\lim_{t\to\infty}S^{cg}(t) = \ln 2$. We compare this result to the numerically estimated $S(t)$, from 
the Langevin equation (\ref{lang}), with the RTP starting at $x=1$, and find that the coarse-grained approximation 
leads to a loss of information (on average) \cite{busiello2019entropy} (see 
Fig.~\ref{fig_entropy}). Notwithstanding this underestimation, the coarse-grained
description does provide a crude measure of the time-scale of relaxation as both the numerical and coarse-grained
estimates seem to agree on the time beyond which they tend to approach their steady state values.
It is to be noted here that unlike the entropy $S(t)$, a numerical estimation of either the entropy 
production $dS_i/dt$ or flow $dS_e/dt$ is a much harder problem due to the presence of the spatial derviatives of the
PDFs $p_\pm(x,t)$.

Interestingly, the fact that both the states $\{+,-\}$ are equally probable, the entropy flow for the coarse-grained
dynamics is identically zero \cite{van2013stochastic}. As a result, the entropy production for the coarse-grained
description equals the total rate of change of entropy, namely:
\begin{align}
\label{dscgi}
\fr{dS^{cg}_i}{dt} = \fr{1}{\tau}(\rho_+(t)-\rho_-(t))\ln\Big(\fr{\rho_+(t)}{\rho_-(t)}\Big) =
\fr{e^{-2t/\tau}}{\tau}\ln\coth\fr{t}{\tau} \geq 0,
\end{align}
with the equality being achieved at the steady state. In addition, the analogy of the relaxation behavior for the
dynamics of the RTP and its coarse-grained description observed in Fig.~\ref{fig_entropy} implies that the RTP relaxes
towards its steady state exponentially (with logarithmic corrections).
It is important, however, to notice a crucial difference between the original dynamics and its coarse-grained 
description.
While the former admits a nonzero entropy production even at the steady state (Eq.~(\ref{dsedt00})), the entropy
production for the latter description identically vanishes at the steady state. Physically, this
happens because the heat bath which is essentially responsible for a nonzero entropy production at the
steady state is now coupled to the dynamical variables, $\rho_\pm(t)$, via the characteristic time $\tau$.

\section{Discussion}
\label{sec_discuss}
\begin{table}[h]
\begin{center}
    \begin{tabular}{ | p{6cm} | l | l |}
    \hline
    Steady state form of PDF & $P_0(x)$ & Eq.~(\ref{p0x_st})\\ \hline
    Laplace form of time-dependent form of PDF & $\tl{P}_0(x,s)$ & Eq.~(\ref{p0x_ss})\\ \hline
    Mean first passage time to cross the barrier starting location $x$ & $\tau(x)$ & Eq.~(\ref{mfpt00})\\ \hline
    Entropy at steady state & $S(t)$ & Eq.~(\ref{ent_ss})\\ \hline
    Entropy production & $dS_i/dt$ & Eq.~(\ref{stsf})\\ \hline
    Entropy flow & $dS_e/dt$ & Eq.~(\ref{stsf})\\ \hline
    Entropy of the coarse-grained dynamics & $S^{cg}(t)$ & Eq.~(\ref{cgst})\\ \hline
    Entropy production of the coarse-grained dynamics & $dS^{cg}_i/dt$ & Eq.~(\ref{dscgi})\\
    \hline
    \end{tabular}
\end{center}
\caption{List of quantities calculated exactly.}
\label{tab1}
\end{table}
In this paper we study the dynamics of run and tumble particles in a bistable potential, in contact with a heat bath. 
Using a piecewise linear function for the potential energy landscape, we map the system of coupled Fokker-Planck
equations describing the dynamics of the RTP to an equivalent matrix differential equation, with constant
coefficients (see Eqs.~(\ref{mat3_ss}),~(\ref{pxs_xgta}), and (\ref{pxs_xlta})).
This mapping allows us to find the exact form of the steady state PDF $P_0(x)$, as well as its time-dependent form
$\tl{P}_0(x,s)$, in Laplace space. Analogous to motion in a monostable potential \cite{singh2025steady}, the 
position
PDF in a bistable potential is also described in terms of a mixture distribution. We find that the time-dependent
PDF relaxes towards its steady state from the center outwards, with the relaxation front moving ballistically
in time. Outside the region in which relaxation has taken place, $P_0(x,t)$ exhibits Gaussian fluctuations
with its amplitude decreasing exponentially in time. We summarize the list of quantities for which we derive
exact forms in Table~\ref{tab1}.

In addition to the transport properties, we also study in this work the escape properties of the RTP from one energy 
minima to another, and using the backward Fokker-Planck equation (\ref{ddqxt}) find the exact
form of the mean first passage time (MFPT) 
to the local energy maxima at the origin. For the interesting case when the RTP starts at the energy 
minima, our analytical estimates of the MFPT agree with the numerically estimated results (within error bounds). The 
confining nature of the potential also implies that first passage time distribution exhibits 
exponential decay, thus allowing us to study the whole process at a coarse-grained level
using a dichotomous Markov process \cite{bena2006dichotomous}.

The advantage of coarse-graining is that it allows us to study the dynamics of an RTP in the bistable potential
as a dichotomous Markov process, a significant simplification of the problem while at the same time retaining the
basic property that the RTP spends maximum amount of time in the energy minima.
However, this advantage comes at a cost that some information about the original dynamics is lost
while performing coarse-graining \cite{busiello2019entropy}. This loss of information is particularly observed while
studying properties of entropy.
We find that while the dynamics of a RTP admits a positive entropy production even at the steady state, the entropy 
production for the equivalent coarse-grained dynamics identically vanishes at the steady state. In addition, while
the original dynamics possesses a nontrivial entropy flow, it vanishes identically under coarse-graining.

The present study raises an important question:
we have assumed that tumbling is a memoryless process, which is more of an idealization for almost all
practical purposes. A particularly important study in this direction would be estimating the characteristic escape
times for a non-Markovian RTP. A phenomena closely related to escape dynamics in a bistable potential is that
of \textit{stochastic resonance}, wherein the transition rates are enhanced under the application of a periodic
signal of appropriate frequency \cite{gammaitoni1998stochastic}. Recent works in Refs.~\cite{qiao2021double}
and \cite{liu2026constructive} studied the properties of stochastic resonance for a Brownian particle, when one varyies the
shape of the underlying bistable potential. An analogous study, involving active particles, would be to investigate the
resonance properties of an RTP in a bistable potential.

We note that, our analysis is based on solving the coupled Fokker-Planck equations describing
the dynamics of an RTP in the piecewise linear symmetric bistable potential. However, a recent work by Frydel \cite{frydel2024run} provides
an alternative viewpoint, in which the continuous trajectories of an RTP are replaced by a jump process, thus allowing to analyze the dynamical properties in terms of integral equations
\cite{frydel2025integral}. 
We could link these works and related studies to ours, in future works.

\section*{Acknowledgments}
RKS acknowledges financial support from Anusandhan National Research Foundation via the Ramanujan Fellowship
project RJF/2026/000006.

\appendix
\section{Roots of cubic equation}
\label{AppR}
Here we look at the conditions which lead to real eigenvalues of the characteristic equation
(\ref{eig3}). Given a cubic polynomial $f(x) = x^3 + a_1x^2 + a_2x + a_3$ with real coefficients $a_1,a_2$ and $a_3$,
then $f(x) = 0$ has three real roots
if its discriminant $\Delta = 18a_1a_2a_3-4a^3_1a_3+a^2_1a^2_2-4a^3_2-27a^2_3 > 0$ \cite{s2004integers}. Now,
the characteristic equation (\ref{eig3}) describing the eigenvalues of the matrix $M_3(b)$ is:
\begin{align*}
 \lambda^3 + \fr{2b}{T}\lambda^2 - \Big(\fr{v^2_0}{T^2} + \fr{2r}{T}
- \fr{b^2}{T^2}\Big)\lambda - \fr{2rb}{T^2} = 0. 
\end{align*}
And the discriminant for the above cubic equation:
\begin{align*}
\Delta = \fr{4v^2_0}{T^6}(v^2_0-b^2)^2+\fr{8rv^2_0}{T^5}(3v^2_0+4b^2)+\fr{4r^2}{T^4}(b^2+12v^2_0)
+\fr{32r^3}{T^3} > 0,
\end{align*}
implying that all the roots of Eq.~(\ref{eig3}) are real.

\section{Details of numerical calculations}
\label{App0}
Here we describe the method we use for numerical calculations in our work.
The Langevin equation (\ref{lang}) describing the dynamics of an RTP in the bistable potential $U(x)$ is
solved using the Euler-Maruyama method:
\begin{align}
x(t+\Delta t)=\begin{cases}
x(t) + (f(x(t)) \pm v_0\Delta t + \sqrt{2T\Delta t}\eta(t),~\text{with probability}~1-r\Delta t,\\
x(t)~\text{and}~v_0 = -v_0,~\text{with probability}~r\Delta t,
\end{cases}
\end{align}
with $r$ being the rate of tumbling, $v_0$ the speed of the RTP, $f(x(t))$ the force acting on the RTP
at position $x(t)$, and $T$ being the temperature.
For the purposes of numerical calculations we use a time step $\Delta t = 0.004$ and for averaging we have
used $10^6$ trajectories.

\bibliographystyle{unsrt}
\bibliography{rtp}

\end{document}